\documentclass[superscriptaddress,showpacs,pre,twocolumn]{revtex4-1}
\usepackage{amsmath,amssymb}
\usepackage{subfigure}
\usepackage{graphicx,bm,mathrsfs,color,hyperref}
\usepackage{ulem,soul}

\begin{document}

\preprint{APS/123-QED}

\title{Chaotic universe model: Lotka-Volterra dynamics of the universe evolution}
\author{Ekrem Aydiner}
\email{ekrem.aydiner@istanbul.edu.tr}
\affiliation{Department of
Physics, Faculty of Science, \.{I}stanbul University,
\.{I}stanbul, 34134, Turkey}%

\date{\today}


\begin{abstract}
In this study, we consider nonlinear interactions between components such as dark energy, dark, matter and radiation in the Friedman-Robertson-Walker space-time framework and propose a simple interaction model based on time evolution of the densities of these components. By using this model we show that these interactions can be given by Lotka-Volterra equation for suitable equation of state parameters. We numerically solve these coupling equations and show that interaction dynamics between dark energy-dark matter-matter or dark energy-dark matter-matter-radiation has a strange attractor for $0>w_{de}>-1$, $w_{dm}\ge 0$, $w_{m}\ge 0$ and $w_{r}\ge 0$ values. These strange attractors with the positive Lyapunov exponent clearly show that chaotic dynamics appears in time evolution of the densities. These results imply that the time evolution of the universe is chaotic in the presence of interactions between these components. The present model may has potential to solve some cosmological problems such as the singularity, cosmic coincidence, crunch, big rip, horizon, oscillations, emergence of galaxies, and large scale organization of the universe. Model also connects between dynamics of the competing species in biological systems and dynamics of the time evolution of the universe, and offers a new perspective and a different scenario for the universe evolution unlike well known popular models.    
\end{abstract}

\maketitle


\section{Introduction}

The formation, structure, dynamics and evolution of the universe has always been of interest. It is commonly accepted that modern cosmology began with the publication of Einstein's seminal article in 1917 \cite{Einstein1917}. Applying the general relativity to the entire universe, Einstein suggested that the universe was static, and spatially curved. Following from this, to explain the structure and dynamics of the universe many interesting models based on Einstein model have been proposed such as flat and expanding universe \cite{deSitter1917}, expanding flat space model, spherical and hyperbolic expanding space \cite{Friedman1922}, original big-bang model \cite{Lemaitre1925,Hubble1926,Lemaitre1927}, expanding flat space \cite{EinsteindeSitter1932}, kinematic expanding models \cite{Milne1933}, oscillating or cyclic universe models \cite{Albrecht1982,Baum2007}, buble universe and inflation bubble universe models \cite{Albrecht1982,Baum2007,Guth1981,Linde1990}, chaotic inflation model \cite{Linde1986} etc. Amongst these models, the big-bang model has been the most accepted one. This is due to the cosmic microwave background (CMB), and cosmic red shift discovered by Hubble observations as well as observations confirming the abundance of light elements in the universe supporting the big-bang scenario. However, new experimental findings such as Type Ia supernovae (SNIa) data \cite{Riess1998,Perlmutter1999,Turner1997,Wang2000}, CMB anisotropy \cite{Spergel2003,Spergel2007}, and large scale structure (LSS) \cite{Tegmark2004,Abazajian2004,Abazajian2005}, showing that the universe does not only expand but does this with an acceleration makes this cosmic scenario more exciting. There is no explanation to this expansion with an acceleration yet. Cosmologists are still working on new models and scenarios to address this situation. One of the best scenarios attempting this is the dark energy. Unfortunately, there is no confirmation of the physical source of this dark energy. Although the origin of the dark energy is not known yet, it is well known that mater is not the only ingredient of the universe. According to what is known today, at present, the universe is composed of approximately $75\%$ dark energy, $20\%$ cold dark matter, $5\%$ baryonic matter and negligible amount of radiation \cite{Riess1998,Perlmutter1999,Turner1997,Wang2000,Spergel2003, Spergel2007,Tegmark2004, Abazajian2004, Abazajian2005}. To explain the nature of the dark energy, there are various dark energy models and mechanisms such as the cosmological constant $\Lambda$ (vacuum or dark energy) proposed. The cosmological constant-cold dark matter-matter model ($\Lambda$CDM) works very well and is in agreement with a large number of recent observations. However, to state without considering highly hypothetical models and their problematic propositions, some of the questions to be answered are the singularity, the cosmic coincidence, the big crunch, the big rip, the horizon, the oscillations, the emergence of galaxies, and the large scale organization of the universe. Despite the great success of the modern cosmology, it is obvious that there can be no success in the development of an integrated theory on the dynamics and evolution of the universe without answering these questions.

There are many proposed models attempting to answer questions above mentioned emerge from the big-bang and other theories based on the theory of the modern cosmology founded by Einstein. For example, oscillating or cyclic universe models, and dark-matter interaction model were proposed to solve the singularity, and fine tuning problems respectively gaining importance in the field. However, no theoretical relationship has been established between the cyclic universe and the dark energy-dark matter interaction models so far, sufficiently addressing the evolution and dynamics of the universe.
Questions on the past and future evolution of the universe, and the mechanisms of the dynamics stemming from this evolution have not sufficiently been addressed. But until now, a more comprehensive scenario has not been developed that addresses the evolution of the universe and the existing cosmological problems. Work based on the interaction of dark energy and dark matter seems to bring optimism to the field \cite{int-Wang2005,int-delCampo2009,int-He2011, Guo2005, Wei2006, Cai2005, Zhang2006, Wu2007, Chen2009, He2009, Chimento2010, Setare2006, Setare2007, Baldi2011, Jian-Hua2008, Khurshudyan2015, Sadeghi2015, Sadeghi2014, Sadeghi2013,int-Pavon2005,int-Wang2005b,int-Feng2007,Miao2004,Ma2010,Ma2009,Xu2013}. This is because the presence of such interactions may have hints that may help to understand the dynamics of the universe leading to a the development of a more realistic scenario. In this work, in contrast to well known popular models, it is assumed that there is a non-linear interaction between dark energy, dark matter, matter and radiation. Although somehow hypothetical, this assumption may have the potential of solving many important problems of cosmology such as the singularity, the cosmic coincidence, the big crunch, the big rip, the horizon, the oscillations, the emergence of galaxies, and the large scale organization of the universe.

The idea of the presence of a non-linear interaction between dark energy, dark matter, matter and radiation may enable the development of a new cosmology scenario on the evolution and the dynamics of the universe. In this work, the interactions between the components forming the universe is modeled and possible outcomes are discussed. This model is a novel one. The interaction models  are based on Friedman-Robertson-Walker (FRW) framework leading to investigation of possible dynamics. We believe in the importance of this work because of the following points: Firstly, the non-linear interaction between the components of the universe were first given by the Lotka-Volterra type equations \cite{Volterra1926, Lotka1925}. It is known that the Lotka-Volterra equation and its variations are mathematical models proposed to model the competition between biological species. It is interesting that the Lotka-Volterra type equation written for cosmology is in the simplest differential form. Secondly, the Lotka-Volterra type equations written for cosmology can give chaotic solutions depending on the values of the parameters of interaction. This is an important outcome for cosmology carrying a potential in helping us to understand questions such as the singularity, the cosmic coincidence, the big crunch, the big rip, the horizon, the oscillations, the emergence of galaxies, the matter distribution and the large scale organization of the universe using the non-linear interaction dynamic. Furthermore, the model proposed here combines the Big Bang and oscillating universe models in a different way and a perspective.

The paper is organized as follows: In the following section to discuss the time evolution of densities, we set interaction equation. We show that coupling equation between dark energy and dark mater can be given by Lotka-Volterra equation. Next, we introduce quadratic interaction which contributes time evolution of the densities. In following section we consider $N$ interacting components to obtain a generalized interaction equation for the universe components. In the same section we numerically analyses time evolution of the densities for $N=3$ (dark energy - dark matter - matter) and $N=4$ (dark energy - dark matter - matter - radiation) in the presence of the interactions. Finally obtained results are discussed.

\section{Coupling interaction between dark energy - dark matter}

A flat Friedman-Robertson-Walker (FRW) space time
whose line elements are given by 
\begin{eqnarray}
ds^{2}=-dt^{2}+a(t)^{2}\overset{3}{\underset{i=1}{\sum}}(dx^{i})^{2}
\label{eq:FRWmetric}
\end{eqnarray}
where $a(t)$ is the scale factor of the three-dimensional flat space, $i$ indicates the spatial components. FRW equations due to the metric~\eqref{eq:FRWmetric} are given by 
\begin{eqnarray}
H^2=\frac{\kappa^2}{3}\rho \ , \qquad \dot{H} = -\frac{\kappa^2}{2}(\rho+p)
\label{eq:FRWequations}
\end{eqnarray}
where $\kappa^{2}=8\pi G$ is the gravitational constant, $H\equiv\frac{\dot{a}}{a}$ is Hubble rate, $\rho$ is energy density and $p$ is pressure. The energy density $\rho$ and pressure $p$ satisfy continuity equation, i.e, energy conservation equation
\begin{eqnarray}
\dot{\rho} + 3H\left(  \rho + p\right) = 0
\label{eq:conservation}
\end{eqnarray}
where over-dot indicates the time derivative. The relation between $\rho$ and $p$ is given by
\begin{eqnarray}
p=w \rho
\label{eq:relation_p_rho}
\end{eqnarray}
where $w$ is EoS parameter which is constant and equal to exactly $-1$ for FRW framework.
For the single fluid the energy conservation~\eqref{eq:conservation} is given in known form 
\begin{equation}
\dot{\rho} + 3H \rho \left( 1 + w\right) = 0 \ .
\label{eq:n-conservation}
\end{equation}

Based on the FRW metric the interaction between dark matter and dark energy can be given as follows:
\begin{subequations}
	\begin{eqnarray}
	\dot{\rho}_{de} + 3H(\rho_{de} + p_{de})=-Q 
	\label{eq:couple-conservation-dark}
	\\
	\dot{\rho}_{dm} + 3H(\rho_{dm} + p_{dm})=Q
	\label{eq:couple-conservation-matter}
	\end{eqnarray}
\end{subequations}
where $Q$ is arbitrary coupling function and subscript stands for a generic dark energy model to be specified \cite{int-Wang2005,int-delCampo2009,int-He2011,Guo2005,Wei2006,Cai2005,Zhang2006,Wu2007,Chen2009,He2009,Chimento2010,Setare2006,Setare2007,Baldi2011,Jian-Hua2008,Khurshudyan2015,Sadeghi2015,Sadeghi2014,Sadeghi2013}. The conservation equations are subject to the Friedman constraint
\begin{eqnarray}
H^2=\frac{\kappa^2}{3}(\rho_{de}+\rho_{dm}) , \   \dot{\rho}{H}=-\frac{\kappa^2}{2}(\rho_{de}+p_{de}+\rho_{dm}+p_{dm}) \ . \quad
\label{eq:no-coupling-FRWequations}
\end{eqnarray}
In this situation, the total energy conservation holds, 
\begin{eqnarray}
\dot{\rho}_{eff}+3H(\rho_{eff}+p_{eff})=0
\label{eq:no-eff}
\end{eqnarray}
where $\rho_{eff}=\rho_{de}+\rho_{dm}$ and $p_{eff}=p_{de}+p_{dm}$ and the FRW  Eq.~\eqref{eq:no-coupling-FRWequations} do change. 

The form of $Q$ is determined under phenomenological assumptions, mainly, the dimensional analysis is used
to construct interactions. It is reasonable to consider interactions which could improve previously known
results and at the same time will not make the mathematical treatment of the problems complicated. It
is widely believed that deeper understanding of the nature of dark energy and dark matter could give
fundamental explanations of the phenomenological assumptions about interaction. There are different $Q$ definition in the literature \cite{int-Wang2005,int-delCampo2009,int-He2011,Guo2005,Wei2006,Cai2005,Zhang2006,Wu2007,Chen2009,He2009,Chimento2010,Setare2006,Setare2007,Baldi2011,Jian-Hua2008,Khurshudyan2015,Sadeghi2015,Sadeghi2014,Sadeghi2013,int-Pavon2005,int-Wang2005b,int-Feng2007,Miao2004}. A simple interaction coupling can be chosen as $Q=\pm \gamma \rho_{dm}\rho_{de}$. In this case, these questions can then be expressed as
\begin{subequations} 	\label{eq:lin-int-dm-de}
	\begin{eqnarray}
	\dot{\rho}_{de} + 3H(\rho_{de} + p_{de})=-\gamma \rho_{dm}\rho_{de} 
	\label{eq:lin-int-de}
	\\
	\dot{\rho}_{dm} + 3H(\rho_{dm} + p_{dm})=\gamma \rho_{dm}\rho_{de}
	\label{eq:lin-int-dm}
	\end{eqnarray}
\end{subequations}
If $\gamma > 0$, the interaction suggests that dark matter is converted into dark energy, while $\gamma < 0$ suggests the inverse process \cite{Mangano2003}. 
By setting $p_{de}=\rho_{de}w_{de}$ and $p_{dm}=\rho_{dm}w_{dm}$
these equations are given by 
\begin{subequations} 	\label{eq:couple-10}
	\begin{eqnarray}
	\frac{d \rho_{de}}{dt} = -3H\left(1+w_{de}\right)\rho_{de} - \gamma \rho_{de}\rho_{dm}  
	\label{eq:couple-con-dark-dif}
	\\
	\frac{d \rho_{dm}}{dt} = \gamma \rho_{dm}\rho_{de} - 3H\left(1+w_{dm}\right)\rho_{dm}
	\label{eq:couple-con-matter-dif} \ .
	\end{eqnarray}
\end{subequations}
Now we can write $r_{1}=-3H\left(1+w_{de}\right) > 0$ for $w_{de}<-1$, and  $r_{2}=3H\left(1+w_{dm}\right)>0$ for $w_{dm}\ge 0$. Hence we obtain 
\begin{subequations} 	\label{eq:couple-11}
	\begin{eqnarray}
	\frac{d \rho_{de}}{dt} = r_{1} \rho_{de} - \gamma \rho_{de}\rho_{dm}  
	\label{eq:a}
	\\
	\frac{d \rho_{dm}}{dt} = \gamma \rho_{dm}\rho_{de} - r_{2} \rho_{dm}
	\label{eq:b} \ .
	\end{eqnarray}
\end{subequations}
By using these relation 
\begin{eqnarray}
x_{1}=\frac{\gamma}{r_{2}}\rho_{de} \ , \quad x_{2}=\frac{\gamma}{r_{1}}\rho_{dm}
\label{eq:transform}
\end{eqnarray}
For a constant or very slowly changing Hubble parameter $H$, Eq.~\eqref{eq:couple-10} with help of Eqs.~\eqref{eq:couple-11} and ~\eqref{eq:transform} can be transformed to
\begin{subequations} \label{eq:LV-dark}
	\begin{eqnarray}
	\frac{dx_{1}}{dt} = r_{1} x_{1}\left(  1-x_{2}\right) 
	\label{eq:LV-dark-dif}
	\\
	\frac{dx_{2}}{dt} = r_{2} x_{2}\left(  x_{1}-1\right) 
	\label{eq:LV-matter-dif}
	\end{eqnarray}
\end{subequations}
where $r_{1} > 0$ for $w_{de}<-1$
and $r_{2} > 0$ for $w_{dm} \ge 0$. 
As it can be seen in Eqs.~\eqref{eq:LV-dark-dif} and~\eqref{eq:LV-matter-dif}, the choice of the interaction term leads us to Lotka-Volterra type equations.  To recall, Lotka-Volterra equations \cite{Volterra1926,Lotka1925} represent the competition between two species and they are used widely in biology, chemistry and various other fields. In this work, the interaction equations between dark energy and dark matter of cosmological systems are used corresponding to the competing prey and predator species in biology.

In this model it is assumed that dark matter has a pressure. Assuming that dark matter has no pressure is purely hypothetical. Furthermore, the assumption of zero pressure contradicts the idea of the change in density. Recently, it is suggested that dark matter has a pressure and various mechanisms have been proposed for this pressure \cite{Harko2012}. In this work, it can be suggested that the dark matter's contribution to pressure may be caused by volume exclusion just like the van der Walls gas. That's why the pressure of dark matter should not be zero. However small, the nonzero value of dark matter means that its EoS parameters should be nonzero too. In other words, for $p_{dm} > 0$ the EoS parameter for dark matter can be given as $w_{dm} > 0$.

Now it is possibly to carry out dynamic and stability analysis of the equations representing the interaction between dark energy and dark matter. The model reaches equilibrium when both of the derivatives are equal to zero.
\begin{subequations}
	\begin{eqnarray}
	r_{1} \left(  1-x_{2}\right) x_{1} = 0
	\\
	r_{2} \left(  x_{1}-1\right) x_{2} = 0 \ .
	\end{eqnarray}
\end{subequations}
When solved for $x_{1}$ and $x_{2}$ the above system of equations yields 
\begin{subequations}
	\begin{eqnarray}
	S_{1}:=\left\{  x_{1},x_{2}\right\}  =\left\{ 0,0 \right\}
	\\
	S_{2}:=\left\{  x_{1},x_{2}\right\}  =\left\{ 1,1 \right\} \ .
	\end{eqnarray}
\end{subequations}
These are fixed points of the coupling equations. 
The stability of the fixed points at the origin can be determined by performing a linearisation by using partial derivation. The Jacobian matrix of the model is
\begin{eqnarray}
J =
\begin{pmatrix}
r_{1} \left(1-x_{2} \right) & - r_{1} x_{1}\\
r_{2} x_{2} & r_{2} \left(x_{1}-1\right)
\end{pmatrix}
\end{eqnarray}
where $J=J\left(x_{1},x_{2}\right)$. When evaluated at the steady state of $\left(0,0\right)$ the Jacobian matrix $J$ becomes
\begin{eqnarray}
J\left(  0,0\right)  =
\begin{pmatrix}
r_{1}   & 0\\
0 & - r_{2}
\end{pmatrix}
\end{eqnarray}
For fixed point $S_{1}=(0,0)$, the eigenvalues of this matrix are $\lambda_{1} = r_{1} = -3H\left(1+w_{de}\right)$ for $w_{de} <-1$
and
$\lambda_{2} = -r_{2} =-3H\left(1+w_{dm}\right)$ for $w_{dm} \ge 0$.
Evaluating $J$ at the second fixed point leads to
\begin{eqnarray}
J\left(  1,1\right)  
=
\begin{pmatrix}
0 & -r_{1}  \\
r_{2}  & 0
\end{pmatrix} \ .
\end{eqnarray}
For fixed point $S_{2}=(1,1)$, the eigenvalues are 
$\lambda_{1}=3iH\sqrt{w_{dm}+1}\sqrt{w_{de}+1}$
and
$\lambda_{2}=-3iH\sqrt{w_{dm}+1}\sqrt{w_{de}+1}$ in the case of $w_{de}<-1$ and $w_{de}\ge 0$. In this model the value of eigenvalues depend on EoS parameters $w_{dm}$ and $w_{de}$. Therefore the characteristic properties of these eigenvalues are determined by sign of EoS parameters. 
The stability of these fixed point are of significance. For the fixed point $S_{1}$ has two real eigenvalues. The relation between eigenvalues is given as $\lambda_{2}>0>\lambda_{1}$ which indicates fixed point $S_{1}$ is a saddle point. 
However, the fixed point $S_{2}$ has two nonzero imaginary eigenvalues as $\lambda_{1,2}=\pm3iH\sqrt{w_{dm}+1}\sqrt{w_{de}+1}$ for $w_{de}<-1$. Hence the linear analysis cannot tell more about nature of $S_{2}$, since the eigenvalues may have null real part. Therefore to understand better how the \textit{limit cycles} behave and on what they do depend on, let's take the ratio of the two equation of the models, hence trying to solve 
\begin{eqnarray}
\frac{dx_{2}}{dx_{1}}=\frac{r_{2} x_{2}\left(  x_{1}-1\right)  }{r_{1} x_{1}\left(  1-x_{2}\right)  }
\label{eq:separate}
\end{eqnarray}
by separating the variables, the equation reads: 
\begin{eqnarray}
\frac{r_{1} \left(  1-x_{2}\right)  }{x_{2}}dx_{2}=\frac{r_{2} \left( x_{1}-1\right)  }{x_{1}}dx_{1}
\end{eqnarray}
integrating now from an arbitrary initial point $\left(  x_{10},x_{20}\right)$ and an arbitrary point $\left(  x,y\right)$, one obtains
\begin{eqnarray}
\Gamma_{1} \left(  \ln x_{2}-x_{2}\right)  + \Gamma_{2} \left(  \ln x_{1}- x_{1}\right)  =F\left(  x_{1},x_{2}\right)
\end{eqnarray}
where $F\left(x_{1},x_{2}\right)$ is the equation of a surface, which depends on the initial conditions $\left(  x_{10},x_{20}\right)$ and represents an invariant of motion. Analyzing Hessian matrix $H(F)$, one can show that $F\left(x_{1},x_{2}\right)$ is a convex function of $\left(x_{1},x_{2}\right)$, that $S_{2}$ represents the critical points, and that the contour lines are close curves. These close curves ones are also the \textit{limit cycles} of the system Eq.~\eqref{eq:LV-dark}, and all the trajectories go onto them. 

It must be noted that for $w_{dm} < -1$ and $w_{de} < -1$ the eigenvalues for $S_{2}=(1,1)$ are be real, hence $S_2$ is the saddle point. For this condition, the trajectory of the dynamic will not be close. This means that, since there will be no cyclic relationship between the two, dark matter and dark energy of the universe will be rapidly reduced to zero or one will be transformed to the other one and disappear completely. Theoretical and observational data indicates that approximately 5\% of the universe is formed of matter and hence any dynamic that causes the disappearance of dark matter cannot be correct. In fact this model establishes lower and upper boundaries for the EoS parameters of dark energy. Assuming that the interaction equations are correct, for a positive and finite value for the dark matter pressure the EoS parameter is given as $w_{dm}>0$, and for negative pressure value of dark energy and cyclic relationships the EoS value should be $w_{de}<-1$. Another point to be noted is the following: Here, the interaction parameter is chosen as $Q = \gamma\rho_{dm}\rho_{de}$. This implies that the transformation between dark energy and dark matter will be equal for both directions. However, a nonlinear transformation approach will be more realistic. The interaction parameters may be chosen different types. This has also been considered in our research but not presented here. The findings show that a different interaction parameter does not affect the characteristic of the solution. This is because of the fact that the interaction parameters lie in the eigenvalues determining the characteristics of the fixed point. 
\begin{figure}[ht]
	\centering
	\includegraphics[width=7cm]{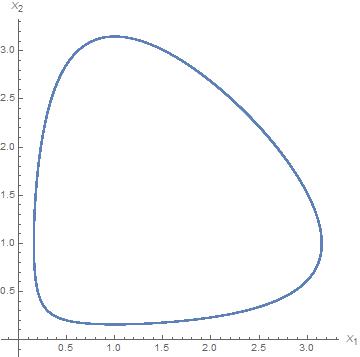}
	\caption{Dimensionless dark matter density $x_{1}$ and dark energy density $x_{2}$ for 1000 time step.}
	\label{fig:fig-cycle}
\end{figure}

A simple numerical solution of Eq.~\eqref{eq:LV-dark} is given in a phase space. Figures~\ref{fig:fig-cycle} shows the interchanges of the densities of dark energy $x_{1}$ and dark matter $x_{2}$ for $r_{1} = 1.0$, $r_{2} = 1.0$. This cycle trajectory is independent of the interaction parameters between dark matter and dark energy but depend on EoS parameters. Cycle solutions are obtained for all $r_{1} > 0$ and $r_{2} > 0$ values. These conditions can be satisfied when taking $w_{de} < -1$ for dark energy and $w_{dm} \ge 0$ for dark matter. Here we assume that the Hubble parameter is a constant for the sake of simplicity. It is well known that the Hubble depends on cosmological time and it takes different value early and late universe era. We consider a special situation where the Hubble parameter $H$ is constant or very slowly changing.

\section{Quadratic interactions in dark sector}

In the previous section by using Eq.~\eqref{eq:relation_p_rho} we set EoS parameters as  $p_{dm}=\rho_{dm}w_{dm}$, $p_{de}=\rho_{de}w_{de}$ for the dark matter and dark energy. In Eq.~\eqref{eq:relation_p_rho} $p=\rho w$ is valid ideal fluid when it is homogeneously distributed in a volume. This means that pressure at the every point of the $V$. Whereas, in realistic systems, pressure may not homogeneously distributed in the volume. In this case pressure $p$ can be written as $p=p(\rho)$ in terms of $\rho$ 
\begin{eqnarray}
p = p(\rho) = \sum_{n=0}^{N}A_{n}\rho^{n} =p_{0}+A_{1}\rho +A_{2}\rho^{2}+... \ .
\label{eq:three-EoS}
\end{eqnarray}
This form of pressure is known as a pertubative pressure or barotropic EoS (See Ref.\cite{Bruni2006}). Here we interested in quadratic form of the EoS $p=p_{0}+A_{1}\rho +A_{2}\rho^{2}$ by ignoring higher order terms. The usual scenario for a cosmological fluid is a standard linear EoS 
($p_{0}=A_{2}=0$), in which case $A_{1}=w$ is usually restricted to the range between $\pm 1$. In high energy regime restricted equation of state can be chosen as $p=A_{1}\rho +A_{2}\rho^{2}$ where the parameter $A_{2}$ set the characteristic energy scale of the quadratic term \cite{Bruni2006}.  

In order to obtain more information about nature of the coupling interactions, equation of states can be modified as $p_{de}=A_{1de}\rho_{de}+A_{2de}\rho_{dm}^{2}$ and $p_{dm}=A_{1dm}\rho_{de}+A_{2dm}\rho_{dm}^{2}$ for inhomogeneous distributed dark energy and dark matter. By setting
$A_{1de}=w_{de}$, $A_{2de}=w_{de}^{\prime}$ and $A_{1dm}=w_{dm}$, $A_{2dm}=w_{dm}^{\prime}$, the Eq.~\eqref{eq:couple-con-matter-dif} and~\eqref{eq:couple-con-dark-dif}  can be reorganized as 
\begin{subequations}
	\begin{eqnarray}
	\frac{d\rho_{de}}{dt}=-3H\left(  1+w_{de}\right)\rho_{de}-3Hw_{de}^{\prime}%
	\rho_{de}^{2} - \gamma\rho_{de}\rho_{dm}
	\label{eq:self-couple-dark-dif}
	\\
	\frac{d\rho_{dm}}{dt} = \gamma\rho_{de}\rho_{dm}-3H\left(  1+w_{dm}\right)
	\rho_{dm}-3Hw_{dm}^{\prime}\rho_{dm}^{2}
	\label{eq:self-couple-matter-dif}
	\end{eqnarray}
\end{subequations}
By using Eq.~\eqref{eq:transform}, these equation can be transformed to
\begin{subequations}
	\begin{eqnarray}
	\frac{dx_{1}}{dt} = r_{1} x_{1} \left(1 - x_{2} \right) - r_{1}^{\prime}x_{1}^{2} 
	\label{eq:x1-self-couple-dark-dif}
	\\
	\frac{dx_{2}}{dt} = r_{2} x_{2} \left(x_{1} - 1 \right) - r_{2}^{\prime}x_{2}^{2}
	\label{eq:x2-self-couple-matter-dif}
	\end{eqnarray}
\end{subequations}
where $r_{1} = -3H\left(1+w_{de}\right)>0$ for $w_{de}<-1$ and $r_{2} = 3H\left(1+w_{dm}\right)>0$ for $w_{dm}\ge 0$ as well in the previous section. Additionally 
$r_{1}^{\prime}=-9H^{2}w_{de}^{\prime}\left(1+w_{dm}\right) \gamma^{-1} > 0$ for  $w_{de}^{\prime}<-1$ and $w_{dm}>0$  values, and $r_{2}^{\prime}=-9H^{2}w_{dm}^{\prime}\left(1+w_{de}\right) \gamma^{-1} > 0$ for $w_{dm}>0$ and $w_{de}<-1$ values.
As can be seen that the quadratic term in rhs of Eq.~\eqref{eq:x1-self-couple-dark-dif} and~\eqref{eq:x2-self-couple-matter-dif} correspond to self-interacting terms in between components of the universe \cite{s-Massey2015,s-Lin2016,s-Nobile2015,s-Boddy2014,s-Mitra2005,s-Bhattacharya2013}. One can clearly see that  these equation are coupled two interacting species such as dark matter and dark energy in the universe and which looks like Lotka-Volterra equations. As a result, by using quadratic EoS we find that coupling interactions lead to self-interacting Lotka-Volterra equations. Our purpose here is to show contribution of the quadratic terms to the dark energy and dark matter interaction. Here we have only studied the dark energy and the dark matter, however, this discussion can be extended to the number of the higher components. 

Now we can briefly discuss the fixed point and stability analysis here to show the dynamics that quadratic contributions will cause. The two equation of the self-interacting models in the phase plane is is given 
\begin{eqnarray}
\frac{dx_{2}}{dx_{1}}=\frac{x_{2}\left( r_{2} + r_{2}^{\prime}x_{2} + r_{2} x_{1}\right)
}{ x_{1}\left( r_{1} + r_{1}^{\prime}x_{1} - r_{1} x_{2}\right)  } \ .
\label{eq:self-x1-stability}
\end{eqnarray}
Unlike Eq.~\eqref{eq:separate}, the phase plane differential equation in Eq.~\eqref{eq:self-x1-stability} is not separable. The simple isoclines are
\begin{subequations}
	\begin{eqnarray}
	\frac{dx_{2}}{dx_{1}}=0\rightarrow r_{1}^{\prime}x_{1} - r_{1} x_{2} = - r_{1}
	\label{eq:xx1-self-couple-dark-dif}
	\\
	\frac{dx_{1}}{dx_{2}}=\infty\rightarrow r_{2} x_{1} + r_{2}^{\prime}x_{2} = - r_{2}
	\label{eq:xx1-self-couple-matter-dif}
	\end{eqnarray}
\end{subequations}
Both of these isoclines are straight lines with positive $x_{1}$ and $x_{2}$ intercepts depending case of i) $r_{1} > r_{1}^{\prime}$
and
$r_{2} < r_{2}^{\prime}$, ii) $r_{1} < r_{1}^{\prime}$
and
$r_{2} > r_{2}^{\prime}$, iii) $r_{1} < r_{1}^{\prime}$
and
$r_{2} < r_{2}^{\prime}$, iv) $r_{1} > r_{1}^{\prime}$
and
$r_{2} > r_{2}^{\prime}$. From time dependent differential equations, one can see that there are two of three equilibrium population of species depending on the EoS parameters in the competing-species model in Eq.~\eqref{eq:x1-self-couple-dark-dif} and~\eqref{eq:x2-self-couple-matter-dif}. In the first and second cases, there are only two equilibrium point which correspond to the extinction of at least one of the species in the universe. However, in the cases of third and fourth there are three equilibrium points in which both species in the universe coexist. This equilibrium population is given by the intersection of the two straight lines $r_{1}^{\prime}x_{1}^{E} - r_{1} x_{2}^{E}= - r_{1}$
and $r_{2} x_{1}^{E} + r_{2}^{\prime}x_{2}^{E} = - r_{2}$. Thus these points are given
\begin{eqnarray}
x_{1}^{E}=-\frac{\left(  r_{1} r_{2} + r_{1}^{\prime} r_{2}^{\prime}\right)  }{\left(
	r_{1} r_{2} + r_{1} r_{2}^{\prime}\right)  }, \quad x_{2}^{E}=-\frac{\left(  r_{1} r_{2} + r_{1}^{\prime} r_{2}\right)  }{\left(  r_{1} r_{2} + r_{1}^{\prime} r_{2}^{\prime}\right)  }
\label{eq:eq-point-1}
\end{eqnarray}
Stability of the coexistent equilibrium population can be discussed for different cases. We can also trajectories of coupling equations~\eqref{eq:x1-self-couple-dark-dif} and~\eqref{eq:x2-self-couple-matter-dif}. However, The trajectory will shift without changing its character.

\section{Generalized equation for nonlinear coupling interactions}

In this section we want to present the generalized equation for coupling interactions.
If we assume that there can be non-linear interactions among multiple components in the universe, then we can write a more general interaction equation. To achieve a more general equation, we start to write the interactions that a single component, such as dark energy, can perform first with itself and with other components, respectively.
\begin{subequations}
	\begin{eqnarray}
	\frac{dx_{1}}{dt} = r x_{1}\left(  1 - r^{\prime} x_{1}\right)  \qquad \qquad \qquad \quad
	\label{eq:more-x0-dark-dif}
	\\
	\frac{dx_{1}}{dt} = r x_{1}\left(  1-\left( r^{\prime}x_{1}%
	+ r^{\prime \prime} x_{2}\right)  \right) \quad
	\label{eq:more-x1-self-couple-dark-dif}
	\\
	\ \frac{dx_{1}}{dt} = r x_{1}\left(  1-\left( r^{\prime} x_{1}%
	+ r^{\prime \prime} x_{2} + r^{\prime \prime \prime} x_{3}\right)  \right)
	\label{eq:more-x2-self-couple-matter-dif} 
	\end{eqnarray}
\end{subequations}
and so on..., where $r$ and $r$ primes are interaction parameters between components. Finally, we can generalize to 
\begin{eqnarray}
\frac{dx_{i}}{dt} = r_{i} x_{i} \left(1 - \sum_{j=1}^{N} \eta_{ij} x_{j} \right) \qquad i=1,...,N
\label{eq:gen-couple-dif}
\end{eqnarray}
where $x_{i}$ denotes the density of the $i$-th species, i.e., $x_{1}$ is the dark energy, $x_{2}$ is the dark matter $x_{3}$ is the matter and $x_{3}$ is the radiation. On the other hand, $r_{i}$ is its intrinsic growth
(or decay) rate and the matrix $\eta_{ij}$ is called the interaction
matrix.  For $i=1$, $x_{1}$ corresponds to dark energy  and rate parameter is $r_{1}=-3H\left(  1+w_{de}\right) $; for $i=2$, $x_{2}$ corresponds to dark matter and its parameter is $r_{2}=-3H\left(  1+w_{dm}\right)$ and so on. On the other hand, matrix elements $\eta_{ij}$ are given by  $\eta_{11}=-\frac{3Hw_{d}^{\prime}\left(  1+w_{m}\right)  }{\eta\left(
	1+w_{d}\right)  }$,  $\eta_{12}=1$,  $\eta_{13}=\frac{\left(  1+w_{3}\right)}{\left(  1+w_{d}\right)}$, and so on. This equation can be called \textit{generalized interacting equation} for all component such as dark energy, dark matter, radiation etc., and this equation likes the equation of competitive species in biological systems (See Ref.\cite{Vano2006,Arneodo1980}). In other words, this equation for $N$ components is the counterpart of the Lotka-Volterra equation in cosmology.

\subsection{Dark energy - dark matter - matter}

\begin{figure}[h!]
	\centering
	\includegraphics[width=7.5cm]{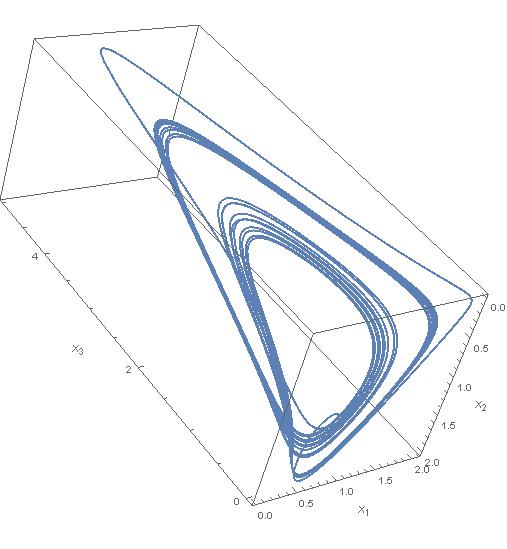}
	\caption{Chaotic attractor for interaction between the dark energy $x_{1}$, dark matter $x_{2}$ and matter $x_{3}$ at $\mu=1.39$ for $w_{de}>-1$, $w_{dm} \ge 0$ and $w_{m} \ge 0$ for 5000 time step.}
	\label{fig:chaotic-atractor-1}
\end{figure}

\begin{figure}[h!]
	\centering
	\includegraphics[width=7.5cm]{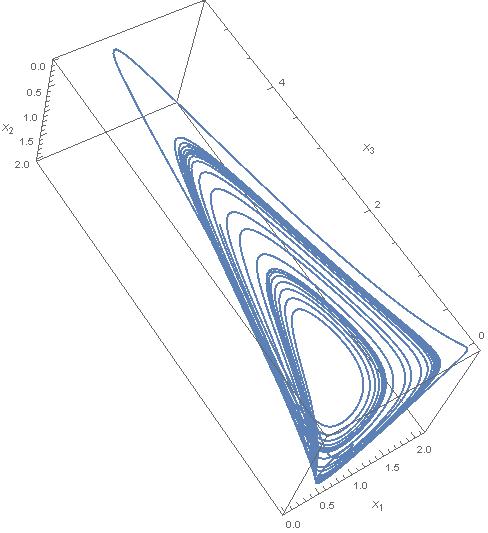}
	\caption{Chaotic attractor for interaction between the dark energy $x_{1}$, dark matter $x_{2}$ and matter $x_{3}$ at $\mu=1.43$ for $w_{de}>-1$, $w_{dm} \ge 0$ and $w_{m} \ge 0$ for 5000 time step.}
	\label{fig:chaotic-atractor-2}
\end{figure}

In this section, we consider coupling interactions among dark energy, dark matter and matter. To obtain numerical for $N=3$ we can write the Eq.~\eqref{eq:gen-couple-dif} in the form \cite{Arneodo1980}
\begin{eqnarray} \label{eq:x0-3coupling}
\frac{dx_{i}}{dt} = x_{i} \sum_{j=1}^{3} \alpha_{ij} (1-x_{j}) 
\end{eqnarray}
where $\alpha_{ij}=r_{i}\eta_{ij}$. Here we consider that EoS for dark energy $w_{de}$ is a little bit greater than -1 which means the dark energy density will slowly decrease as the universe expand. In order to numerically solve the Eq.~\eqref{eq:x0-3coupling} we chose $0>w_{de}>-1$, $w_{dm} \ge 0$ and $w_{m} \ge 0$ that provides the conditions $r_{1}<0$, $r_{2}>0$ and $r_{3}>0$. The parameters $\alpha_{ij}$ to solve Eq.~\eqref{eq:x0-3coupling} are given by 
\begin{eqnarray} \label{eq:x-data-set}
\alpha_{ij}=
\begin{pmatrix}
-0.5 & -0.1 & 0.1 \\
0.5 & 0.5 & 0.1 \\
\mu & 0.1 & 0.1
\end{pmatrix}
\end{eqnarray}
For these $\alpha_{ij}$ parameters satisfy $r_{1}=-0.5$, $r_{2}=1.1$ and $r_{3}=\mu+0.2$. By using data set ~\eqref{eq:x-data-set} we obtain phase space solutions for two different arbitrary $\mu>0$ values. Obtained numerical results for $\mu=1.39$ and $\mu=1.43$ are given in Figures~\ref{fig:chaotic-atractor-1} and \ref{fig:chaotic-atractor-2}, respectively. In Figure~\ref{fig:chaotic-atractor-1} we set initial values for $\mu=1.39$ at $t=0$ we set $x_{1}=0.2$, $x_{2}=0.3$ and $x_{3}=0.14$. As it can be clearly seen from  Figure~\ref{fig:chaotic-atractor-1} for these initial conditions and data set given in ~\eqref{eq:x-data-set} Eq.~\eqref{eq:x0-3coupling} has a strange attractor. By using TISEAN package program we compute the largest Lyapunov exponent as $\lambda=0.045$. On the other hand in Figure~\ref{fig:chaotic-atractor-2} for $\mu=1.43$ at $t=0$ we set $x_{1}=0.3$, $x_{2}=0.7$ and $x_{3}=0.1$.  Figure~\ref{fig:chaotic-atractor-2} also shows that dynamics of Eq.~\eqref{eq:x0-3coupling} has a chaotic attractor with Lyapunov exponent $\lambda=0.091$. These strange attractors provide that the dynamics of coupling interactions between dark energy, dark matter and matter is chaotic. These solutions are repeated for different initial and different data set provides the conditions $0>w_{de}>-1$, $w_{dm} \ge 0$ and $w_{m} \ge 0$ and find similar chaotic behavior in the coupling dynamics.   

In this section we have shown that, unlike binary interactions, triple coupling interactions lead to chaotic dynamics. This is, of course, is a very simple model and is based entirely on the idea that the constituents of the universe are transformed to each other through interactions. This simple model, inspired by the competition of species in biological systems, offers remarkable results on the universe dynamics.    

Here obtained results shows that chaotic behavior appears in coupling interaction dynamics when dark energy parameter takes $0>w_{de}>-1$ values. In cosmology many experimental observations provides that dark energy EoS parameter $w_{de}$ bigger than minus one for different era of the universe. Therefore, these observations provide $0>w_{de}>-1$ supports the idea that the universe may have chaotic behavior in any period. It is well known that if a physical system has a chaotic dynamics it creates an order according to the chaotic dynamics and does not easily leave itself in order. The importance of chaotic dynamics for the universe discuss in next section.

If we go back the biological systems, we know that the competing systems for $N=3$ admits limit cycles behavior.
Vano et al. \cite{Vano2006} studied the occurrence of chaos in basic Lotka-Volterra models of Lotka-Volterra models of
four competing species. Apparently, for $N\leq3$ chaos is not possible. However for $N=3$ it is shown that chaotic behavior can appear in Eq.~\eqref{eq:x0-3coupling} for unphysical parameters \cite{Arneodo1980}. Contrary to $N=3$ Lotka-Volterra equations our equation in Eq.~\eqref{eq:x0-3coupling} shows that the chaotic behavior appears in universe due to interactions between dark energy, dark matter and matter for realistic parameters. 

\subsection{Dark energy - dark matter - matter - radiation}

\begin{figure}[h!]
	\centering
	\includegraphics[width=7.5cm]{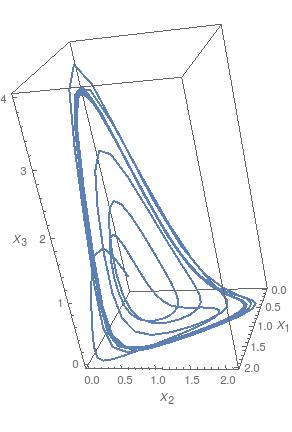}
	\caption{Chaotic attractor for interaction between the dark energy $x_{1}$, dark matter $x_{2}$, matter $x_{3}$ and radiation $x_{4}$ for 2000 time steps.}
	\label{fig:chaotic-d4}
\end{figure}

Here we also consider quadrapole coupling interaction between dark energy, dark matter, matter and radiation densities. For these interactions we set $0>w_{de}>-1$, $w_{dm} \ge 0$, $w_{m} \ge 0$ and $w_{r} \ge 0$. To obtain numerical for $N=3$ we can write the Eq.~\eqref{eq:gen-couple-dif} in the form \cite{Arneodo1980}
\begin{eqnarray} \label{eq:x0-4coupling}
\frac{dx_{i}}{dt} = x_{i} \sum_{j=1}^{4} \alpha_{ij} (1-x_{j}) 
\end{eqnarray}
where $\alpha_{ij}=r_{i}\eta_{ij}$.
The parameters $\alpha_{ij}$ to solve Eq.~\eqref{eq:x0-4coupling} are given by
\begin{eqnarray} \label{eq:x-data-d4-set}
\alpha_{ij}=
\begin{pmatrix}
-0.5 & -0.1 & 0.1 & 0.1\\
-0.5 & 0.5 & 0.1 & 0.2\\
0.3 & 0.1 & 0.1 & -0.3 \\
0.1 & 0.1 & 0.1 & 0.1
\end{pmatrix}
\end{eqnarray}
For these $\alpha_{ij}$ parameters satisfy $r_{1}=-0.6$, $r_{2}=0.6$, $r_{3}=0.2$ and $x_{4}=0.4$.  By using initial conditions $x_{1}(0)= 0.21$, $x_{2}(0)= 0.35$, $x_{3}(0)= 0.11$ and $x_{4}(0)= 0.51$ at $t=0$, Eq.~\eqref{eq:x0-4coupling} for four species is solved. Obtained results is shown in Figure~\ref{fig:chaotic-d4}. We also find positive Lyapunov exponent $\lambda=0.011$ for these data. It can be clearly shown that coupling interactions for arbitrary but $0>w_{de}>-1$, $w_{dm} \ge 0$, $w_{m} \ge 0$ values and $w_{r} \ge 0$ and arbitrary $\alpha_{ij}$ values has a chaotic solutions. Here these solutions are also repeated for different initial and different data set provides the conditions $0>w_{de}>-1$, $w_{dm} \ge 0$ and $w_{m} \ge 0$ and find similar chaotic behavior in the coupling dynamics.   

\section{Concluding Remarks}

In this study, it is taken into account that components such as dark energy, dark matter, matter and radiation interacted each other in the universe, and the time-dependent behavior of densities of these components are examined. Several steps are followed in the study. \textit{In the first step}, we consider only the interaction between dark energy and dark matter, and we show that this interaction can be represented by an equation of Lotka-Volterra type given in Eqs.~\eqref{eq:LV-dark-dif} and~\eqref{eq:LV-matter-dif} for the conditions $w_{de}<-1$, $w_{dm}\ge 0$ and $w_{m}\ge 0$. Results in Figure~\ref{fig:fig-cycle} show that there is a limit cycle behavior between dark energy and dark matter for $w_{de}<-1$, $w_{dm}\ge 0$ and $w_{m}\ge 0$. \textit{In the second step}, we consider the quadratic contribution to the coupling interactions for the  conditions $w_{de}<-1$, $w_{dm}\ge 0$ and $w_{m}\ge 0$ and we show that these interactions can be given by Eq.~\eqref{eq:x1-self-couple-dark-dif} and~\eqref{eq:x2-self-couple-matter-dif}. \textit{In the third step}, we consider $N$ coupling interactions between components of the universe. We generalized these interaction equations the help of quadratic interactions for arbitrary interaction parameters. We show that the generalized equation can be given in the form of ~\eqref{eq:gen-couple-dif} which is similar to the generalized Lotka-Volterra equation \cite{Vano2006,Arneodo1980}.
\textit{In the fourth step} we consider triple coupling interactions between dark energy, dark matter and matter and we solve numerically Eq.~\eqref{eq:x0-3coupling} for the data set in  ~\eqref{eq:x-data-set} for two different arbitrary $\mu>0$ values. Obtained numerical results are shown in  Figures~\ref{fig:chaotic-atractor-1} and \ref{fig:chaotic-atractor-2} that chaotic behavior appears in the dynamics of the coupling interactions for the conditions $0>w_{de}>-1$, and $w_{dm}\ge 0$ and $w_{m}\ge 0$ values. \textit{In last step}, we study coupling interactions between different four species such as dark energy, dark matter, matter and radiation. We numerically analyses  Eq.~\eqref{eq:x0-4coupling} for four species by using data in ~\eqref{eq:x-data-d4-set}. We find in Figure~\ref{fig:chaotic-d4} that for $N=4$ interactions also has chaotic behavior for the conditions $0>w_{de}>-1$, $w_{dm} \ge 0$ and $w_{m} \ge 0$ and arbitrary $\alpha_{ij}$ parameters. Discussions were made for a special situation where the Hubble parameter has a constant or very slowly changing with time. In the last two step EoS parameter for the dark energy is set as $w_{de}>-1$ which imply that dark energy density will slowly decreases while the universe expanding. The EoS parameters used in this study have been chosen as consistent with literature 
\cite{Riess1998,Perlmutter1999,Turner1997,Wang2000,Spergel2003, Spergel2007,Tegmark2004, Abazajian2004,Abazajian2005,int-Wang2005,int-delCampo2009,int-He2011,Guo2005,Wei2006,Cai2005,Zhang2006,Wu2007,Chen2009,He2009,Chimento2010,Setare2006,Setare2007,Baldi2011,Jian-Hua2008,Khurshudyan2015,Sadeghi2015,Sadeghi2014,Sadeghi2013,int-Pavon2005,int-Wang2005b,int-Feng2007,Miao2004,Ma2010,Ma2009,Xu2013} and experimental analyses in Refs. \cite{Riess1998,Perlmutter1999,Turner1997,Wang2000,Spergel2003, Spergel2007,Tegmark2004, Abazajian2004,Abazajian2005}.

The model proposed here is based on interactions between components of the universe and dynamics is modeled over the density evolution of these components. The present model shows that the universe evolution is chaotic at least for the some special parameter values in the presence of the interaction between components. This model and its possible results may present a new understanding and perspective on the dynamics and evolution of the universe. We summarize the possible results of the chaotic universe dynamics: 1) Chaotic dynamics presents a very different universe scenario from big-bang. If the universe is evolving due to such a dynamics, it means that the universe oscillates without repeating itself. Such a scenario has not singularity, big crunch or big rip. When the dark energy density increases, the universe begins to expand. However when the dark energy begins to turn into dark matter or matter, the gravitational force becomes dominant and the universe shrinks again due to gravitational force. In any case, according to this chaotic dynamics, the process continues with oscillations that do not repeat themselves. 2) This scenario can also contribute to making the cosmic coincidence problem understandable. 3) It may be possible to explain the horizon problem with regard to the emergence of similar universe formations in regions that can not communicate with each other, since this model allows the local interactions between dark energy, dark matter, matter and matter. Even if there is no interaction with each other, similar interaction dynamics can lead to the similar formations in the area outside the horizon. 4) Local interactions can also provide a mechanism for the galaxy formation. 5) On the other hand, if the universe has a chaotic dynamics, it can be explained the occurrence of the fractal patterns from the micro-cosmos to the macro-cosmos at all scales. 6) This model allows us to write a very simple form for dynamical of the universe.

Finally we note that the model connects between dynamics of the competing species in biological systems and dynamics of the time evolution of the universe, and offers a new perspective and a different scenario for the universe dynamics and evolution unlike well known popular models, even though it needs more experimental confirmation.

\section*{Acknowledgements}

In this study, Mathematica package program is used to obtain the numerical solutions and plotting. The largest Lyapunov exponent is computed with help of the TISEAN package program. Author grateful to Alper Tunga Aydiner for useful discussions. Author also thanks Ferhat Nutku, Erkan Yilmaz and Orhan Gemikonakli for kind help.

\end{document}